# High-speed proton therapy within a short breath-hold

*Vivek Maradia[1\*], Nick Yue[2], Adam Molzahn[2], Jingqian Wang[2], Mark Pankuch[3],*

*Serdar Charyyev[1\*], Billy W. Loo Jr.[1,4\*]*

[1]*Department of Radiation Oncology, Stanford University School of Medicine, Stanford, California, USA*
[2]*Mevion Medical Systems, Littleton, Massachusetts, USA*
[3]*Department of Medical Physics, Northwestern Medicine Proton Center, Warrenville, Illinois, USA*
[4]*Stanford Cancer Institute, Stanford University School of Medicine, Stanford, California, USA*

*\*Corresponding authors*

**Co-senior authors**:

Serdar Charyyev, PhD (charyyev@stanford.edu)

Billy W Loo Jr, MD PhD (BWLoo@Stanford.edu)

**Author for editorial correspondence**:

Vivek Maradia, PhD (maradia@stanford.edu)

**Acknowledgements:**

V.M. acknowledges financial support from the Berry Foundation. We gratefully acknowledge the discussions and managerial support provided by the leadership, research, and engineering teams at Mevion Medical Systems, including Tina Yu, Mark Jones, Skip Rosenthal, Lionel Bouchet, Townsend Zwart, and Ying Xiong.

**Disclosures:**

NY, AM, and JW are employed by Mevion Medical Systems. BWL is co-founder and board member of TibaRay and has received lecture honoraria from Mevion Medical Systems.

**Authors contribution:**

VM designed and conducted the study and drafted the initial manuscript. NY performed the treatment planning. AM and JW assisted with the treatment time calculations. MP provided the upright CT data for the study and input on gantry-less treatment workflow. SC and BWL critically revised the manuscript at all stages and supervised the study through regular meetings. All authors have read, revised, and approved the final manuscript.

**Abstract:**

Proton therapy provides superior dose conformity compared with photon radiotherapy, concentrating radiation within the tumor while sparing adjacent healthy tissue. This advantage has been most effectively realized for static tumors in anatomically stable regions, such as the head and neck. For thoracic and abdominal sites, however, physiological motion remains a critical challenge: because the proton dose distribution is highly sensitive to density variations, long delivery times relative to respiratory motion can compromise accuracy. Existing strategies to accelerate delivery often require substantial hardware modifications or are difficult to translate into routine practice.

Here we report an optimization that enables high-speed proton delivery (5-10 sec per field) on a commercial synchrocyclotron platform without hardware changes. The method combines high-energy shoot-through beams with Bragg-peak delivery, an optimized nearest-neighbor scanning sequence, and a two-pulse dose regulation scheme. Applied to eight lung cancer cases (target volumes 100–1000 cc), the approach achieved full-field delivery in under 10 s—compatible with a short breath-hold—while preserving conformity, dose accuracy, and sparing of organs at risk.

This framework provides a practical route to motion-robust proton therapy, improving precision, efficiency and patient tolerance. More broadly, it opens a pathway toward widespread clinical adoption of high-speed proton delivery for moving tumors.

**Main:**

Proton therapy is an emerging modality for localized cancer treatment that exploits the unique ballistic properties of protons to deposit the majority of radiation dose at a sharply defined depth within tissue—the Bragg peak (BP)[1,2]. Intensity-modulated proton therapy (IMPT) dynamically scans multiple BPs with energy layer specific patterns across the tumor volume[1], enabling highly conformal dose distributions that spare surrounding healthy tissue.

Although numerous simulation studies have demonstrated the dosimetric advantages of IMPT over photon-based radiotherapy[1,2] (i.e., using high-energy x-rays), definitive clinical evidence of improved patient outcomes is still emerging. Early clinical data, however, indicate benefits such as reduced acute toxicities and improved overall survival in head and neck cancer[3,4], improved overall survival in patients with leptomeningeal metastases from breast and lung cancer[5], fewer late side effects in pediatric brain tumor survivors[6], and diminished rates of treatment-related sequelae including lymphopenia, secondary malignancies, and impaired quality of life[7–12].

Despite these promising indications, proton therapy faces two major obstacles to widespread adoption: high infrastructure costs and technical factors constraining the full potential benefits of proton therapy[13–15]. As Bortfeld and Loeffler have argued[15], technological solutions can make proton therapy more comparable to photon treatments in size and cost. A promising strategy to achieve this involves transitioning to gantry-less systems, which drastically reduce facility size, complexity, and expense. Recent commercial innovations exemplify this trend: the synchrotron-based P-Cure system[16] and the synchro-cyclotron based MEVION S250-FIT system[17], currently being commissioned at the Stanford Medicine Cancer Center, both fit within vaults comparable in size to conventional photon linac vaults, offering scalable and more cost-effective proton therapy solutions.

While these compact systems reduce capital costs, pencil beam scanning (PBS) remains highly sensitive to anatomical changes and motion during prolonged beam delivery, particularly for tumors that undergo breathing-induced movement such as lung and upper abdomen tumors[18]. Common motion mitigation strategies—including breath-hold[19–22], rescanning[23–28], gating[29–31], or their combination—address the interplay effect between tumor motion and beam delivery, but often prolong overall delivery time. For example, gating treatments for large lung tumors can extend to 45 minutes or more per session, compromising patient comfort or tolerance as well as clinical throughput and cost-effectiveness[32,33].

To enhance access and reduce costs, increasing patient throughput while effectively managing tumor motion is essential[34]. Delivering treatment under quasi-static conditions using breath-hold techniques—in which patients hold their breath during irradiation to minimize motion—would be most effective if an entire treatment field can be delivered within a short breath hold of 5 to 10 seconds, achievable by nearly all patients even with compromised lung function. In PBS proton therapy, total treatment time depends on both beam-on time and dead time, the latter being the interval needed to switch energy layers and the time to change lateral spot positions.

Some institutions have previously proposed various techniques to achieve ultra-fast treatment delivery in proton therapy systems equipped with gantries [34–42]. However, these methods require extensive modifications to the beamline and treatment planning system, posing significant challenges for broad clinical implementation. In this work, we aim to establish a framework for high-speed treatment delivery on the compact, gantry-less MEVION S250-FIT commercial proton therapy system by developing simple and easily implementable methods that require no hardware modifications. These techniques are designed to enable short breath-hold treatments and improve the clinical feasibility of proton therapy for moving lung tumors. Our approach is expected to reduce field delivery times to 5–10 seconds for a wide range of tumor volumes (100–1000 cc), making rapid and precise treatment accessible within existing clinical infrastructures.

An additional pathway to further reduce treatment costs is accelerated dose intensification, which decreases the total number of fractions by delivering higher doses per session. The safety of such accelerated regimens depends upon limiting the dose to normal organs. An exemplary goal is to achieve delivery of 6 Gy (RBE) per fraction (rather than a conventional 2 Gy (RBE) per fraction) within a short breath-hold per treatment field, or potentially even an entire treatment session within a single breath hold. Combining high-dose per session with robust motion mitigation through high-speed delivery promises to improve patient comfort, increase throughput, and substantially enhance the cost-effectiveness and clinical impact of proton therapy.

### High-speed approach to lateral penumbra sharpening in PBS proton therapy:

Cyclotron- and synchrocyclotron-based proton therapy systems typically deliver fixed-energy beams (230–250 MeV)[43]. To treat patients, however, a variable energy range (typically 70–230 MeV) is required. The energy selection is achieved by using an energy degrader, which unfortunately increases the beam size, emittance, and energy/momentum spread[37]. To mitigate these effects, a collimator system is positioned after the degrader, followed by an energy selection system (ESS) to reduce the beam size and restore desired energy properties[39].

In contrast, MEVION S250-FIT systems are compact and do not include a traditional beamline. Instead, they use a set of 12 range shifter plates of different thicknesses placed just before the patient to achieve the necessary clinical beam energies. However, this approach increases the beam size—typically by a factor of 2–3 compared to other commercial systems—resulting in broader lateral penumbra and increased dose to healthy tissue surrounding the tumor[44].

To address this, an adaptive aperture (AA) system, similar in concept to multi-leaf collimators (MLCs) in photon therapy, shapes the beam at the tumor periphery for each energy layer, improving the lateral dose fall-off and enhancing plan quality (Figure 1(b))[45,46]. However, the use of AA introduces time delays, adding approximately 5–10 seconds to the field delivery time, depending on tumor size and shape. To achieve high-speed field delivery (within 5–10 seconds), it becomes essential to develop alternative strategies for sharpening the lateral penumbra without incurring time penalties.

One effective strategy to achieve sharper lateral dose fall-off in proton therapy is to vary the beam size across the tumor volume—using smaller beam sizes at the periphery and larger sizes in the central region. This concept was first introduced by Maradia et al. demonstrating the potential for improved dose conformity by tailoring beam optics to the anatomical structure of the target[41]. However, in practice, dynamically altering the beam size during delivery is challenging, as most systems do not support on-the-fly spot size modulation without either collimation or significant time penalties.

To overcome this limitation, we propose a solution that exploits the physical properties of high-energy proton beams. Specifically, the highest available beam energy (*e.g.*, 230 MeV) can be used in a "shoot-through" (ST) configuration for the outer tumor regions. These high-energy beams have smaller spot sizes and undergo less lateral scattering within the patient, enabling sharper dose gradients. For the interior of the tumor, conventional BP beams with larger spot sizes can be used to maintain dosimetric efficiency and depth conformity. This combined strategy—using ST beams at the edges and BPs centrally—offers a practical and time-efficient method to enhance lateral penumbra without modifying hardware or delivery time (as shown in Figure 1). Our objective is to substitute AA with ST beams while maintaining a comparable dose distribution.

Other recent studies have highlighted potential benefits of using ST beams. Combining ST beams with rotational arc delivery could improve plan conformity for some of the most difficult targets in base-of-skull tumors[47,48]. This approach makes a trade-off between better high dose conformity from the sharper lateral penumbra of ST beams combined with arc delivery *vs*. the depth dose advantage and lower integral dose of BP beams, which are considered the traditional strength of proton therapy. Kong et al. investigated the use of ST beams in combination with BP for head-and-neck cancer with the focus on improving dose to nearby organs at risk (OAR). Their automatic planning study demonstrated sharper lateral penumbra and improved OAR sparing compared to conventional IMPT[49]. These studies were not focused on increasing delivery speed, which would take on the order of a few minutes on conventional proton treatment systems. Of note, in gantry-based systems, to switch between transporting high-energy ST beams and lower energy BP beams through the bends of the beamline, the beamline magnets must be ramped from low-to-high or high-to-low energy, which takes several seconds (approximately 10 seconds or longer)[50].

Our proposed approach on a gantry-free platform eliminates the time penalty of switching between ST and BP beams. Additionally, the use of ST beams will require a beam stop for the beam transmitted through the patient. This is greatly simplified in the gantry-free design because the beam orientation is in a single fixed direction.

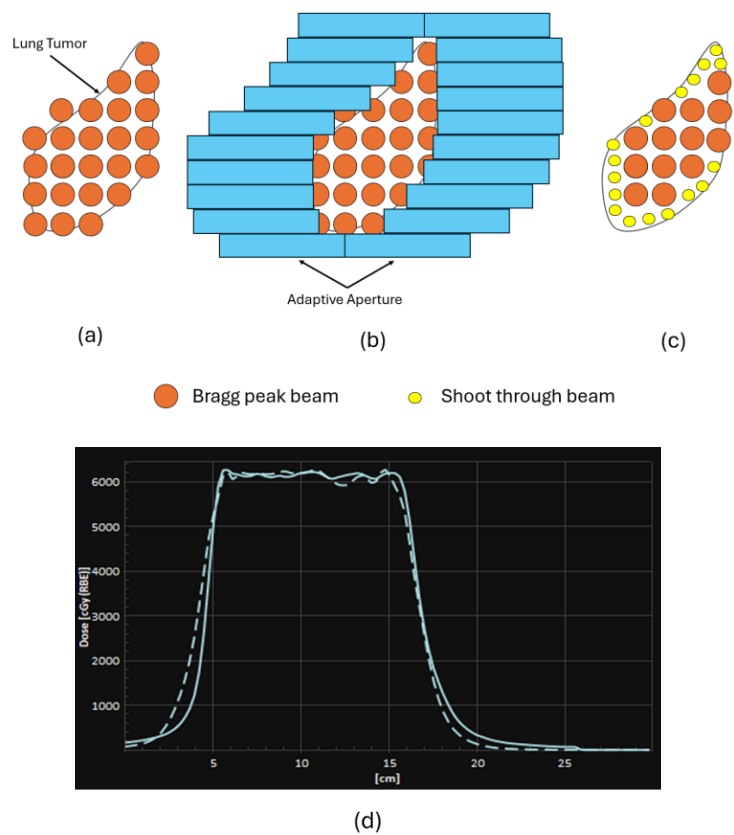

*Figure 1 Schematic representation of energy layer from a lung cancer treatment plan illustrating spot placement for different delivery techniques: (a) Standard BP plan; (b) BP plan with adaptive aperture to sharpen the lateral dose fall-off at the tumor edge; (c) Combined approach using smaller ST beam spots at the tumor periphery and larger BP spots at the center to enhance lateral dose conformity. (d) Dose profiles showing sharpness of lateral dose fall-off is comparable between ST beam (solid line) versus AA (dashed line) (example from plan P1).*

We produced treatment plans for eight non-small cell lung cancer cases, varying in tumor size and shape, using our novel approach that integrates ST beams with BP delivery. These plans were benchmarked against conventional plans generated using AA to ensure comparable clinical quality. The plans using the combined technique demonstrated similar or improved target coverage and organ-at-risk sparing without compromising dose conformity or homogeneity. Table 1 and Table 2 summarizes key parameters for each plan, including the number of fields used, the number of energy layers, the total number of spots, and dose metrics within the planning target volume (PTV) as well as doses received by critical OARs surrounding the tumor. These results demonstrate that our method that achieves lateral penumbra sharpening using ST beams rather than collimation with the adaptive aperture maintains clinical plan quality with faster delivery and without requiring hardware modifications. A comparison of treatment plans using the ST and BP technique and those generated with the AA approach is presented in Table 2.

*Table 1 Overview of target and plan characteristics across different non-small cell lung cancer cases, including the size of the planning target volume (PTV), and number of fields. Energy layers and scanning spots per field are reported for the combination of ST and BP plans.*

| Plan | PTV (cc) | Field | Number of energy Layers | Number of spots |
|---|---|---|---|---|
| P1 | 102 | F1 | 9 | 612 |
|  |  | F2 | 9 | 618 |
| P2 | 220 | F1 | 13 | 957 |
|  |  | F2 | 15 | 602 |
|  |  | F3 | 12 | 307 |
| P3 | 294 | F1 | 15 | 1157 |
|  |  | F2 | 14 | 1318 |
|  |  | F3 | 13 | 1503 |
| P4 | 432 | F1 | 16 | 1217 |
|  |  | F2 | 15 | 873 |
|  |  | F3 | 15 | 837 |
| P5 | 602 | F1 | 15 | 2590 |
|  |  | F2 | 13 | 2693 |
| P6 | 671 | F1 | 17 | 2381 |
|  |  | F2 | 13 | 2685 |
| P7 | 876 | F1 | 17 | 3352 |
|  |  | F2 | 18 | 2451 |
| P8 | 1046 | F1 | 14 | 2924 |
|  |  | F2 | 18 | 3463 |

Table 2 Overview of dose parameters for ST+BP plans without AA and BP plans with AA to the PTV, along with doses to organs at risk (OAR). Dose–volume metrics are defined as follows: $V_{95}$ = percentage of the volume receiving ≥95% of the prescribed dose; $D_2$ = dose received by the hottest 2% of the volume.

| Plan | PTV dose parameters | | | | | | Dose in OAR (mean dose in %) | | | | | |
|---|---|---|---|---|---|---|---|---|---|---|---|---|
| | Mean dose (%) (ST+BP) | Mean dose (%) (BP+AA) | $V_{95\%}$ (%) (ST+BP) | $V_{95\%}$ (%) (BP+AA) | $D_{2\%}$ (%) (ST+BP) | $D_{2\%}$ (%) (BP+AA) | Dose in Lung (ST+BP) | Dose in Lung (BP+AA) | Dose in Heart (ST+BP) | Dose in Heart (BP+AA) | Dose in spinal cord (ST+BP) | Dose in spinal cord (BP+AA) |
| P1 | 100.7 | 100.4 | 94 | 94.1 | 103.4 | 103.5 | 6.65 | 6 | 13.7 | 13.9 | 13.1 | 15 |
| P2 | 101 | 100.9 | 90.6 | 90.5 | 104.2 | 104 | 15 | 14.9 | 2.7 | 3 | 3.9 | 4.5 |
| P3 | 101.6 | 101.7 | 92.1 | 92 | 104.7 | 104.9 | 35.5 | 34 | 9.5 | 9.7 | 9.8 | 10.2 |
| P4 | 101.2 | 101.3 | 93 | 93.1 | 103.6 | 103.3 | 12.1 | 11.1 | 3.2 | 3.4 | 4.1 | 4.8 |
| P5 | 101.5 | 101.2 | 94 | 93.8 | 103.7 | 103.5 | 3.6 | 3.5 | 14.2 | 15.1 | 11.8 | 11.9 |
| P6 | 100.7 | 100.3 | 92.3 | 92.1 | 103.3 | 103.2 | 23.5 | 22.7 | 30.8 | 31.2 | 1.1 | 1.1 |
| P7 | 102.1 | 101.8 | 93.2 | 93.1 | 104.7 | 104.5 | 33.6 | 32 | 54.9 | 56 | 17.1 | 17.9 |
| P8 | 100.6 | 100.4 | 90.7 | 90.5 | 104.2 | 104.1 | 15.4 | 15 | 0.1 | 0.4 | 16.4 | 17.2 |

### Optimized scanning sequencing to minimize spot scanning time:

Conventional proton therapy systems deliver dose using a line-by-line scanning sequence. Within each energy layer, the proton beam moves laterally from one spot to the next in a row-wise fashion before stepping to the next line, pausing briefly between positions. The time required for each spot transition depends on the scanning magnet speed, and when treating large or complex tumors, the cumulative time spent on lateral movements can become a significant portion of total delivery time.

The MEVION S250-FIT system, based on a synchrocyclotron, delivers pulsed proton beams with a repetition rate of 1.3 ms. Its scanning magnets operate at a speed of 6 mm/ms, allowing a maximum lateral spot displacement of 7.8 mm during each 1.3 ms pulse-off window. If the distance between two consecutive spots is ≤7.8 mm, the repositioning occurs entirely within this window, effectively resulting in zero additional scanning time. However, conventional line-by-line scanning often results in many inter-spot distances exceeding this threshold, incurring time penalties.

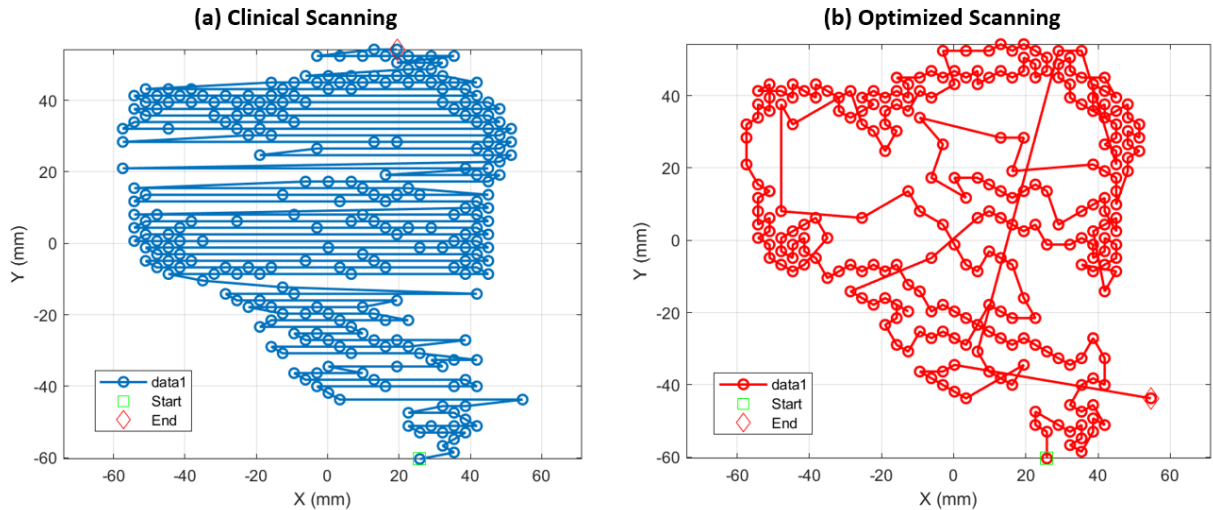

*Figure 2 Comparison between (a) conventional clinical scanning sequence using line-by-line spot delivery and (b) optimized scanning sequence based on a nearest-neighbor algorithm to reduce scanning time. In this example, the layer scanning time can be reduced from 455 ms to 65 ms with optimization.*

To address this, we propose a novel spot sequencing approach using a nearest-neighbor algorithm that dynamically selects the next closest spot within the same energy layer, regardless of line order. For example (as shown in Figure 2), in a typical 126.6 MeV energy layer with 361 total spots, conventional scanning results in 108 transitions exceeding 7.8 mm, leading to a scanning time of ~455 ms. In contrast, our nearest-neighbor optimized pattern reduces this to just 24 long-distance transitions, cutting scanning time to ~65 ms for the same energy layer. This optimization dramatically reduces lateral positioning time without hardware changes, enabling faster field delivery.

Alternative scanning techniques have been explored to increase voxel-based dose rates in order to exploit the potential biological advantages of FLASH proton therapy [51,52]. On the other hand, the focus of the current work is a different optimization to minimize the overall field delivery time.

**Energy change time:**

In the MEVION S250-FIT system, energy changes are achieved by adjusting the range shifter plates rather than altering the energy by using an energy degrader at the cyclotron or synchrotron source. This design eliminates the need for complex magnetic field adjustments or energy selection system (ESS) typically required in conventional proton therapy systems. As a result, energy layer switching can be achieved in only 50 milliseconds compared to energy layer switching times ranging from 80 milliseconds to 2 seconds in other typical designs[53].

**Optimizing charge delivery time/beam on time:**

The Mevion machine utilizes range shifters for energy modulation, resulting in negligible beam losses except for nuclear losses in the degrader. This design enables the Mevion system to deliver very high beam currents directly at the patient location. Because there are no significant beam losses along the beamline, the synchrocyclotron provides clinically used beam current of 1-2 nA for patient treatment, which is sufficient to deliver the required dose to each spot with a single pulse. Consequently, one spot can be treated within approximately 1.3 ms.

However, the Mevion synchrocyclotron's cold cathode ion source inherently exhibits instability, causing roughly ±20% variation in beam current for a fixed pulse width. To ensure precise dose delivery at each spot, the current clinical system employs a multi-step feedback process during the final pulse sequence. It begins by delivering 70% of the target charge, measures the actual dose delivered, and then adjusts subsequent pulses to deliver 75% of the remaining charge, followed by 80%, and finally 100% of what remains. This stepwise adjustment compensates for beam current fluctuations and maintains dose accuracy within ±3%. Yet, this approach requires four pulses per spot, each lasting about 1.3 ms, totaling approximately 5.2 ms per spot. For a treatment field with 3000 spots, this adds up to roughly 15.6 seconds to complete the dose delivery.

To improve overall treatment speed, a two-pulse feedback scheme is proposed. In this method, the system first delivers around 85% of the target charge in the initial pulse and measures the actual delivered dose. It then adjusts a single second pulse to deliver the remaining 15%. By concentrating most of the dose delivery in the first pulse and requiring only one correction pulse, this approach simplifies the regulation process and significantly reduces delivery time. Since the remaining charge to be corrected is relatively small, the impact of beam current fluctuations on final dose accuracy is minimized. To evaluate dose delivery accuracy using a two-pulse feedback scheme, we simulated 10 million spots and found that dose accuracy within a ±2% (FWHM) tolerance could be achieved. The histogram of dose errors for individual spots is shown in Figure 3. With this streamlined method, delivering dose to 3000 spots would take only about 7.8 seconds, enhancing treatment efficiency without sacrificing safety or precision.

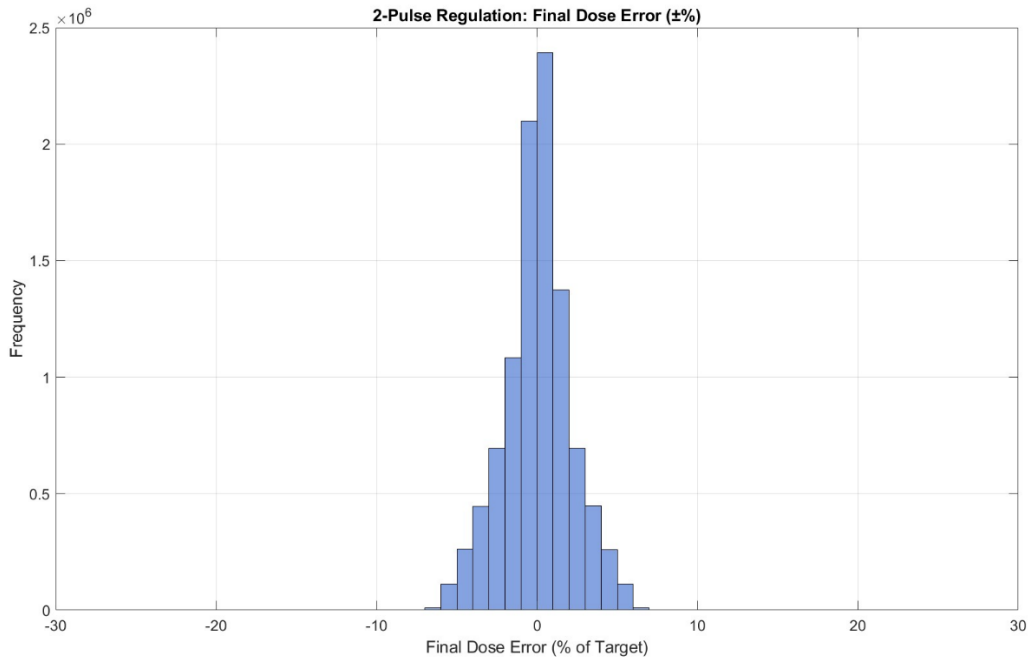

*Figure 3 Histogram showing the distribution of dose deviation (%) during the simulated delivery of 10 million spots using a 2-pulse feedback approach, demonstrating dose accuracy within a ±2% (FWHM) tolerance.*

**Assessing total field delivery time:**

Although the MEVION S250-FIT system at our facility is not yet clinically operational, delivery time estimates were obtained by analyzing treatment delivery log files from the Mevion production site in Littleton using demonstration runs. As shown in Figure 4 (a), the field delivery times for all plans were under 10 seconds. The total energy change time across all plans was consistently less than 1 second, and the total scanning time per field was also under 1.5 second. The primary contributor to the overall delivery time was the delivery charge, accounting for approximately 80% of the total time. Nonetheless, even the charge delivery time remained below 8 seconds for all cases, regardless of tumor size. Figure 4(a) illustrates detailed delivery time breakdowns for each field across all eight lung tumor plans. These results confirm the technical feasibility of sub-10-second field delivery, with clinical implementation planned upon system commissioning at Stanford. Figure 4(b) shows the reduction in total delivery time with our proposed method compared to the conventional setting (BP with AA), assuming no tumor motion (recognizing that this is a poor assumption when delivery times are long). Delivery times were reduced by over 90% across all fields, enabling short breath-hold treatment without any mechanical modifications to the machine.

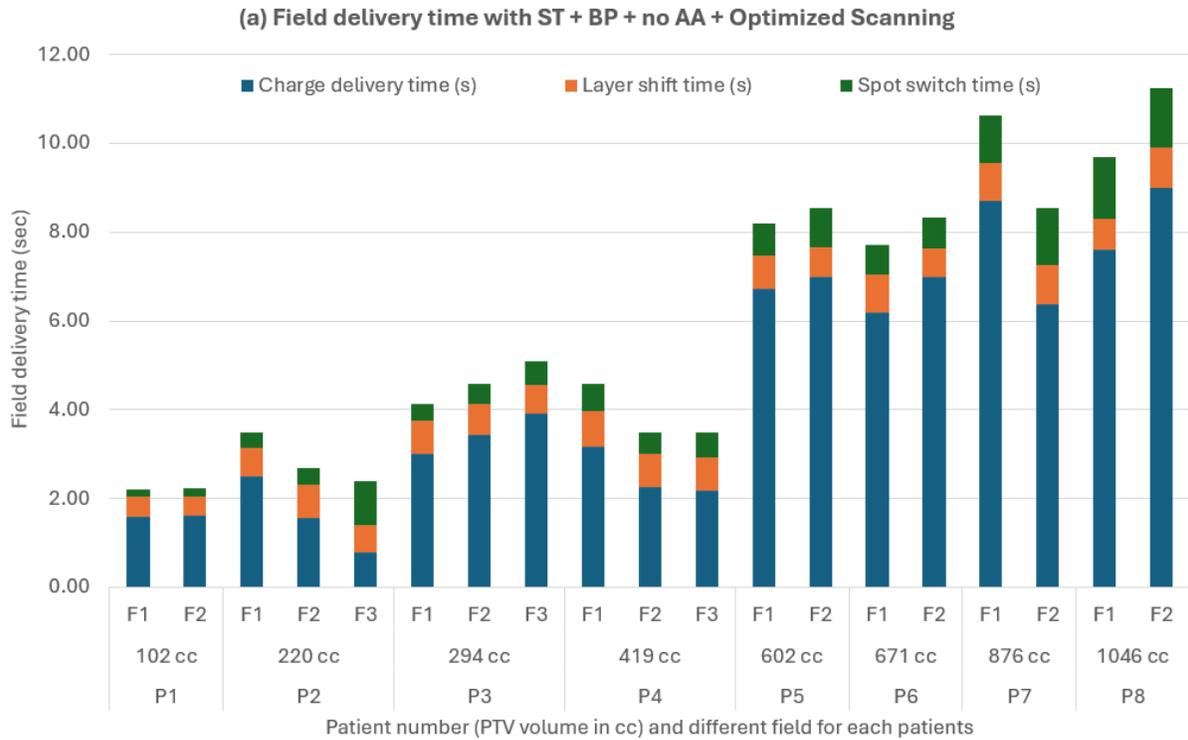

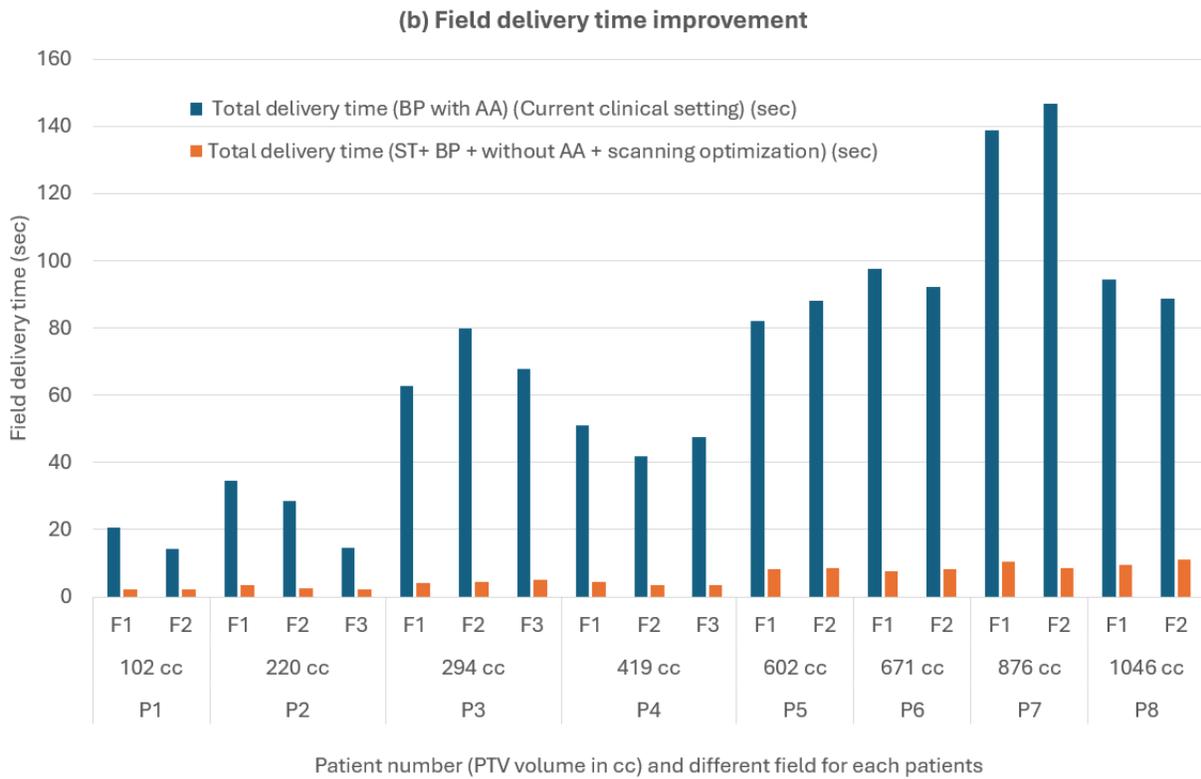

*Figure 4 (a) Field delivery times for eight lung cancer cases using ST beams with BP, without AA, and with optimized scanning. (b) Total delivery time comparison between clinical plans (BP with AA) and ST plans (BP without AA, optimized scanning), assuming no tumor motion.*

Of note, our approach is scalable to a broad range of dose per session without increasing delivery time. The beam current from the synchrocyclotron can be increased in a straightforward way from a typical 1-2 nA to 3-5 nA. The two-pulse feedback scheme enables each spot to be treated in the same duration, resulting in no additional time needed for higher-dose treatments. Furthermore, the Mevion synchrocyclotron is capable of delivering beam currents up to 25 nA (with improvement in the ion source it can potentially deliver more than 50 nA beam current), providing significant flexibility and capacity for future dose escalation strategies such as stereotactic ablative radiotherapy with doses exceeding 10 Gy per session while maintaining fast and precise treatment delivery.

**Practical realization and clinical relevance:**

Conventional motion management strategies in radiotherapy, particularly scanning pencil beam proton therapy, are inefficient and potentially error prone. For example, respiratory gating, in which the beam is synchronized to be delivered only during a specified portion of the breathing cycle, prolongs the overall treatment time because of its limited duty cycle and is subject to residual error because the delivery spans multiple non-identical breathing cycles. Because each field can take multiple minutes to deliver, there is a greater need to repeat imaging verification between fields to address potential anatomical drifts that can occur in a patient over a prolonged treatment session, which further adds to the overall treatment time. Such treatments for large lung tumors can take 45 minutes to an hour per session, inclusive of setup and imaging verification in addition to the beam delivery itself. If each field could be treated in a short breath hold of 5-10 seconds, the motion-induced dose uncertainty and overall treatment time could be reduced dramatically through compounding effects, as shorter treatment facilitates greater patient comfort and reduced fatigue, leading to improved positional stability and reduced need for repeated imaging verification during the treatment session. Total treatment session times could be reduced potentially to 15 minutes or less, enhancing clinical efficiency.

We have shown that by combining a novel lateral dose fall-off sharpening technique—utilizing high-energy ST beams at the tumor periphery and BPs at the core—with optimized scanning patterns and efficient dose feedback, it is possible to treat even large lung tumors with high doses within a short breath-hold of 5 to 10 seconds during a given treatment field. Many patients, even with impaired lung function, can maintain longer breath holds of more than 30 seconds, or even longer with supportive strategies such as pre-oxygenation. This opens the possibility of completing an entire multi-field delivery in a single breath hold.

This approach promises to address some of the main challenges in proton therapy, including sensitivity of dosimetric accuracy to physiologic motion and cost effectiveness limited by low clinical throughput in capital intensive facilities. Importantly, it requires no hardware modifications to the MEVION S250-FIT system, enabling straightforward clinical translation. It is also compatible with Mevion gantry-based systems employing the same beamline design and therefore applicable to a larger installed base. Furthermore, all synchrocyclotron-based proton therapy systems, including many existing installations from other manufacturers, can benefit from our solution. Although these systems typically have relatively long energy switching time (~800 ms per step), our improvements in the scanning and beam on time components can still achieve treatment delivery times of 15–30 seconds per field, depending on tumor size, a significant improvement over current performance of several minutes per field.

Our approach has some additional advantages and potential for future development. Using an adaptive aperture to sharpen the beam and improve lateral dose fall-off results in neutron production through interaction of the proton beam with the aperture material, which contributes a small but unwanted integral dose to the patient. By combining ST beams at the tumor edges with BP delivery at the center, the need for an aperture is eliminated—thereby avoiding any aperture-induced neutron dose to the patient.

The use of ST beams with a gantry-free fixed beamline architecture enables straightforward integration of a range telescope on the beam exit side of the patient (in place of a beam stop)[48]. This would allow real-time, high-precision measurements of the residual energy of protons after passing through tissue, providing accurate verification of proton range and dose delivery, and improving safety and quality assurance by identifying energy deviations or setup errors. It also supports continuous monitoring of dose and dose rate throughout treatment—all without requiring any hardware modifications to the treatment machine itself.

Tomographic imaging of positron emitting radionuclides generated by the interaction of the proton beam with tissue has been proposed as a method of verifying delivered dose and proton beam range. High-speed proton delivery may particularly facilitate this prompt PET imaging by increasing the signal from short half-life isotopes, supporting real-time adaptive strategies and enhancing confidence in treatment accuracy, particularly for complex or mobile tumors.

In treatment of pediatric cancers, one of the best recognized applications of proton therapy, fast delivery provides a major clinical advantage by reducing or eliminating the need for anesthesia. Children often require general anesthesia to comply with strict immobilization during longer treatments, adding risk and complexity. By delivering the dose in just a few seconds, the need for anesthesia is markedly reduced. Combining high-speed treatment with strategies such as the AVATAR system for audio-visual distraction[54,55] supports awake, stable patient positioning, and can dramatically reduce the use of invasive immobilization and anesthesia procedures.

Finally, rotational arc proton delivery has been proposed as a promising approach to overcome dose distribution challenges in complex tumors[56]. However, current delivery times are often several minutes per field, and this is not a commonly used technique. Applying our strategy on the MEVION S250-FIT system is expected to reduce arc delivery times to under 40-50 seconds, making arc proton therapy clinically practical.

Our optimization of proton therapy delivery provides a path toward making high-speed, motion-robust, and cost-effective proton therapy clinically practical. By leveraging existing commercial hardware with minimal system adjustments, this approach offers a scalable solution that can be rapidly implemented across proton therapy centers. The ability to treat large tumors within a single breath-hold, support dose intensification, and reduce reliance on anesthesia in pediatric settings represents an important andvance in addressing the technical and logistical barriers limiting proton therapy's broader adoption.

## Methods:

### Treatment Planning Strategy:

Eight non-small cell lung cancer (NSCLC) cases were selected to evaluate a novel approach to high-speed proton delivery. For each case, two distinct treatment plans were generated using a research version of the RayStation treatment planning system (RaySearch Laboratories, Stockholm, Sweden) equipped with a validated beam model of the MEVION S250-FIT synchrocyclotron system. This system uses a binary range shifter mechanism integrated into the nozzle for energy modulation, yielding nominal proton energies between 12 and 230 MeV.

The two planning approaches were:

1. **Bragg Peak with Adaptive Aperture:** Traditional plans incorporating layer-specific collimation to sharpen lateral dose fall-off using the dynamic AA system.

2. **Combined Bragg Peak and Shoot-Through:** Collimator-free plans enhanced with an additional high-energy (230 MeV) ST layer. These plans used spot filtering to fully populate the target projection—including margins—with 230 MeV spots. During optimization, target-specific dose objectives guided the selection and weighting of these high-energy spots to improve peripheral dose conformity.

All plans were optimized using the RayStation 2024B Monte Carlo dose engine with a constant RBE of 1.1.

### Beam Configuration and Selection:

Beam angles were carefully chosen based on patient-specific anatomical considerations to ensure robust plan quality and minimize sensitivity to setup and range uncertainties. Preferred beam orientations minimized path length through regions of heterogeneous density (*e.g.*, ribcage) and avoided traversing organs at risk (OARs) such as the heart. A library of beam configurations was developed based on tumor location (*e.g.*, left *vs.* right lung, superior *vs.* inferior lesions), with the goal of minimizing beam range.

### Robust Optimization Framework:

All treatment plans were generated using 3D robust multi-field optimization. A comprehensive uncertainty model was applied, incorporating 5 mm setup error and 3% range uncertainty, resulting in 21 perturbed scenarios for each plan. Target volumes (CTVs) were robustly optimized across all scenarios, while dose constraints for OARs were applied in the nominal scenario only.

Optimization objectives included dose coverage constraints (minimum and maximum dose) for the PTVs, mean dose constraints for the lungs (excluding GTVs) and heart, and maximum dose limits for the spinal cord and surrounding healthy tissue. Dose fall-off was enforced via constraints on a body contour structure. Objective weights were tuned manually for each plan, typically ranging from:

- **CTV objectives**: 500–1000
- **Body dose fall-off**: 1000
- **Spinal cord**: 50
- **Lung and heart**: ~1

Additional constraints and structures were added as needed to account for individual anatomical variations and plan quality.

### MEVION S250-FIT Beamline/Nozzle description:

A schematic of the MEVION S250-FIT beamline is shown in the figure below. The system features a synchrocyclotron accelerator that delivers a fixed 230 MeV pencil beam at the exit port. The beamline includes an air-core, iron-free scanning magnet capable of steering the beam in both X and Y directions, maintaining a constant source-to-axis distance (SAD) and beam divergence. Six transmission ion chambers (TICs) are integrated into the nozzle—four for strip detection in both transverse planes and two for integral charge measurement—providing real-time feedback on beam position, shape, and dose.

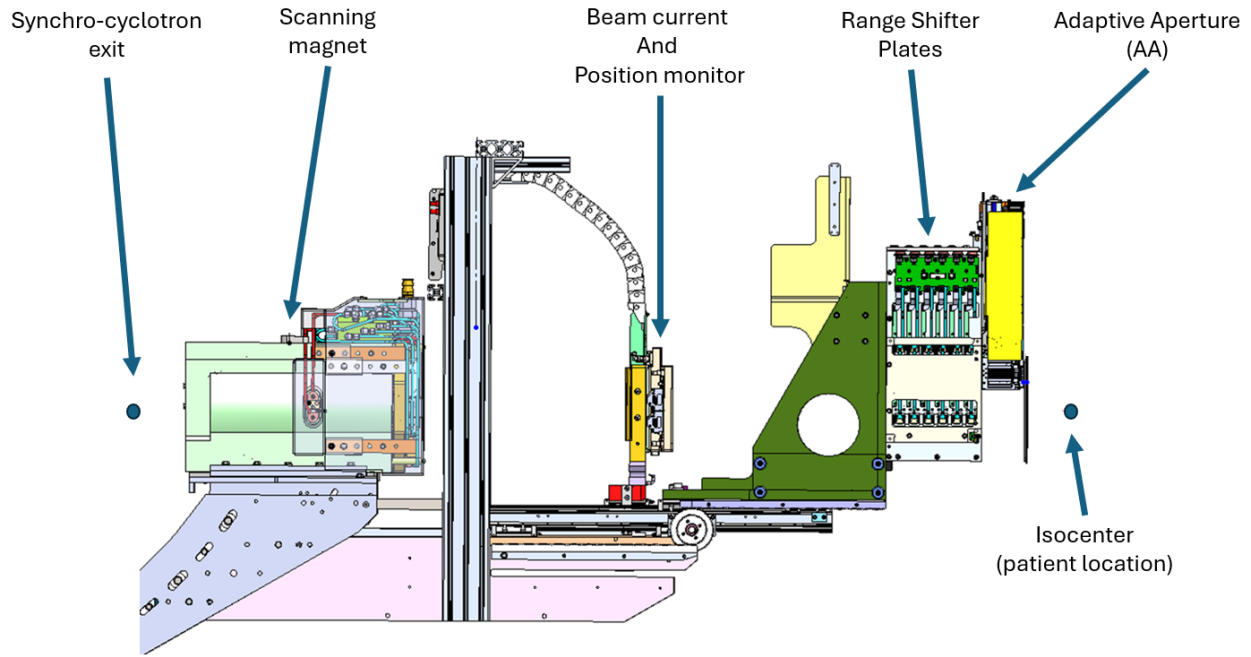

*Figure 5 The MEVION S250-FIT beam line schematic in the maximum extension configuration (30cm total extension range).*

Fast energy modulation is achieved through a range shifter composed of 12 plates: 10 boron carbide and 2 Lexan, enabling precise range control from surface to 32.2 cm water-equivalent depth in 0.25 g/cm² steps. The plate arrangement supports modular range shifting, with combinations of "Oct," "Quad," "Double," and "Single" boron carbide plates used for stepwise modulation, while the upstream Lexan plates ("Half" and "Quarter") fine-tune the overall range. The downstream positioning of thicker plates minimizes beam scattering, and the plate thicknesses are optimized around BP spacing. This system enables fast, precise energy layer modulation for highly conformal, layer-by-layer dose delivery.

At the end of the beamline, a dynamic field collimation system simulates patient-specific apertures by trimming or blocking individual pencil beams across the treatment field. Unlike conventional collimators, this system can shape each energy layer independently and adjusts its position relative to the isocenter to minimize the air gap. The collimator defines the nozzle-to-isocenter distance, which can vary from 8.6 cm to 38.6 cm. Both the collimation system and the range shifter are mounted on an extension mechanism, allowing for real-time adjustment to maintain optimal geometry and reduce lateral penumbra. This integrated design enables highly conformal, layer-specific beam shaping with rapid energy modulation—essential for ultra-fast and precise dose delivery in advanced proton therapy.

**Treatment delivery time calculation:**

As the MEVION S250-FIT system at our site is still in the commissioning phase, treatment delivery times were not measured on-site. Instead, we obtained detailed system performance data—including energy layer switching time, spot-to-spot transition time, and beam-on (charge delivery) time—from demonstration log files recorded at the Mevion production facility in Littleton, Massachusetts. These data, collected during controlled demonstration runs, were used to assess the timing characteristics of each component contributing to total field delivery. This analysis enabled accurate breakdown of the delivery process and verification of the system's capability to achieve high-speed treatment within the targeted 5–10 second time window. Clinical implementation and in-patient timing validation will follow system commissioning.

**Nearest Neighbor Algorithm for Optimized Spot Scanning:**

In conventional pencil beam scanning proton therapy, lateral dose delivery within each energy layer typically follows a line-by-line or raster scanning pattern. While simple to implement, this approach can be inefficient—especially for large or irregularly shaped fields—because it results in long lateral movements between distant spots, increasing the total scanning time.

To address this, we implemented a nearest neighbor (NN) algorithm to optimize the spot delivery sequence within each energy layer. Rather than scanning in a fixed line order, the NN algorithm dynamically selects the next spot based on proximity: after delivering a spot, the system identifies and moves to the closest undelivered spot. This process continues recursively until all spots in the layer are delivered.

By minimizing the distance between successive spot positions, the NN algorithm substantially reduces the number of long magnet transitions—particularly those that exceed the maximum repositioning range allowed during the synchrocyclotron's 1.3 ms pulse-off interval. As a result, many spot movements can be completed without incurring additional scanning delays.

In our tests, this method reduced lateral scanning time by more than 80% compared to the conventional pattern for the same number of spots, without changing the dose distribution. This makes it particularly well-suited for high-speed treatment delivery.